\documentstyle[12pt,aasms4]{article}

\slugcomment{To be published in {\it The Astrophysical Journal}}
\def\sun{\ifmmode\odot\else$\odot$\fi}

\def\degns{\ifmmode^\circ\else$^\circ$\fi}
\def\deg{\ifmmode^\circ\else$^\circ$\fi\ }

\def\h2co{\hbox{H$_2$CO}}

\def\HII{\hbox{H\hskip 2trueptII }}
\def\H76{\hbox{H~76$\alpha$}}
\def\H2O{\hbox{H$_2$O}}
\def\h2o{\hbox{H$_2$O}}
\def\degns{\ifmmode^\circ\else$^\circ$\fi}
\def\deg{\ifmmode^\circ\else$^\circ$\fi\ }

\def\arcs{\char'175~}

\def\arcm{\char'023~}
\def\kms{~km~s$^{-1}$}
\def\kmss{~km~s$^{-1}$~}

\def\eg{e.g.,~}

\def\etal{et~al. }

\def\NH3{\hbox{NH$_3$}}

\def\wff{W\thinspace{44} }
\def\wffns{W\thinspace{44}}
\def\wte{W\thinspace{28} }
\def\wtens{W\thinspace{28}}

\begin{document}

\title{THE SIZES OF 1720 MHZ OH MASERS:  VLBA AND MERLIN OBSERVATIONS 
OF THE SUPERNOVA REMNANTS \wff AND \wte}

\author{M. J Claussen, W. M. Goss, D. A. Frail, \& K. Desai}
\affil{National Radio Astronomy Observatory (NRAO)
Array Operations Center, P.O. Box O, Socorro, New Mexico 87801, USA}
 


\begin{abstract}

We have used the NRAO Very Long Baseline Array (VLBA) to image OH(1720
MHz) masers in the supernova remnants W28 and W44 at a resolution of
40 mas. We also used MERLIN to observe the same OH(1720 MHz) masers in
W44 at a resolution of 290 $\times$ 165 mas. All the masers are resolved 
by these VLBA and MERLIN
observations. The measured sizes range from 50 to 180 mas and yield
brightness temperature estimates from 0.3-20$\times{10}^8$ K. 
We investigate whether these measured angular sizes are intrinsic and 
hence originate as a result of the physical conditions in the supernova 
remnant shock, or whether they are scatter broadened sizes produced by 
the turbulent ionized gas along the line of sight. While the current 
data on the temporal and angular broadening of pulsars, masers and 
extragalactic sources toward W44 and W28 can be understood in terms of 
scattering, we cannot rule out that these large sizes are intrinsic.  
Recent theoretical modeling by Lockett \etal suggests that the physical
parameters in the shocked region are indicative of densities and OH abundances
which lead to estimates of sizes as large as what we measure.
If the sizes and structure are intrinsic, then the OH(1720 MHz) masers 
may be more like the OH(1612 MHz) masers in circumstellar shells than 
OH masers associated with \HII regions.  At two locations in W28 we observe 
the classical S-shapes in the Stokes {\it V} profiles caused by Zeeman 
splitting and use it to infer magnetic fields of order 2 milliGauss.  

\end{abstract}


\section{Introduction}

The 1720.53 MHz line from the hydroxyl radical (OH) was conclusively
shown to be associated with the supernova remnant (SNR)
W\thinspace{28} by Frail, Goss \& Slysh (1994). Since then several
surveys have been made toward other Galactic SNRs (Frail et al. 1996,
Yusef-Zadef et al. 1996, Green et al. 1997, Koralesky et al. 1998),
clearly establishing that the OH(1720 MHz) line toward masers is a new
class of OH maser, distinct from those in star-forming regions and
evolved stars. Follow-up work (Claussen et al. 1997, Frail \& Mitchell
1998, Wardle, Yusef-Zadeh \& Geballe 1998) supports the hypothesis
that the OH(1720 MHz) masers originate in C-type shocks, transverse to
the line-of-sight, being driven into adjacent molecular clouds by the
expanding SNR. The measured densities, temperatures and magnetic
fields from these studies are consistent with collisional excitation
of the OH by the H$_2$ molecules in the post-shock gas (Elitzur 1976,
Pavlakis \& Kylafis 1996, Lockett, Gauthier, \& Elitzur 1999).

%
Claussen \etal (1997) imaged the masers toward the SNRs \wte and \wff 
and reported finding numerous maser features distributed across these 
SNRs.  At their arcsecond resolution, some features were unresolved 
(``spots'') while other features appeared to be resolved with measured 
angular sizes ranging from 0.25'' to 2.5''.  
These measured  sizes of the masers could, of course,  be spots that 
appeared spatially 
blended due to insufficient angular
resolution.  Alternatively, the resolved features could be spots whose
apparent sizes reflect scattering by interstellar turbulence along the
line of sight (angular broadening).  Many masers are
known whose apparent sizes are dominated by angular broadening (\eg 
Diamond \etal 1998; Frail \etal 1994); most such masers are seen with 
similar sizes and elongations as noted by Claussen \etal (1997).

The current study was undertaken primarily to address the question of 
whether the measured size of the masers is due to scattering or to 
multiple maser components.  In order to reach a definite conclusion
on this question, we must investigate both the intrinsic size of the
masers and the possible effects of interstellar scattering. 

The OH(1720 MHz) masers are quite rare in the interstellar medium as 
compared to main-line masers, and even more so toward late-type stars 
where the other satellite line at 1612 MHz is the dominant maser 
transition.  So published sizes of OH(1720 MHz) masers, which necessitate 
the use of VLBI techniques are also very rare. Forster \etal (1982) 
found upper limits of 20 mas to the sizes of OH(1720 MHz) masers toward 
the \HII region NGC~7538, while Masheder \etal (1994) reported that the 
W3(OH) 1720 MHz masers are unresolved with sizes $<$1.2 mas.  These 
measurements may not even be applicable in the case of supernova 
remnants where the physical conditions and pumping mechanisms could be 
very different from that in star-forming regions.  A recent study of 
the pumping of the 1720 MHz masers toward supernova remnants 
(Lockett \etal 1999) finds tight constraints on the physical conditions 
needed for their production (temperature in the range 50 --- 125 K, 
molecular hydrogen density $\sim$ 10$^5$ cm$^{-3}$, and OH column 
densities of order 10$^{16}$ cm$^{-2}$.)  An upper limit for the size 
of the maser spots is the thickness of the shocked region over which 
such conditions exist.  This thickness is estimated to be about
3$\times$10$^{15}$ cm.  At the distance (3 kpc) of both \wff and \wte, 
this corresponds to an angular size of about 60 mas.  

The effects of interstellar scattering on the sizes of masers in \wte
and \wff can be constrained by observations of nearby (in projection)
pulsars and extragalactic sources.  Interstellar scattering of an
extragalactic source results in angular broadening while interstellar
scattering of pulsars results in pulse broadening, an increase in the
apparent width of a pulsar's average pulse profile beyond its
intrinsic width.  The degree of angular broadening for the masers and
the extragalactic source and the degree of pulse broadening for the
pulsar all depend in different ways upon the relative geometry of the
observer, scattering material, and sources.  Measurements
of these three scattering effects can constrain the distribution of
the scattering material and can be used to estimate, for example, the
unscattered sizes of the masers.

In this paper, we present VLBI observations of several 
bright OH(1720 MHz) masers in \wte and \wff using the VLBA of the 
NRAO. The large extent of the SNRs and associated maser emission 
(tens of arcminutes) precluded observations of all the OH masers for 
these remnants.  In addition, we have used the MERLIN telescope of the
Nuffield Radio Astronomy Laboratory at Jodrell Bank to 
observe the OH masers in \wff with intermediate angular resolution 
between that of the VLBA and the VLA.  We present the results of these 
observations and discuss their impact on the question of the masers' 
intrinsic size vs. broadening due to interstellar scattering.

\section{Observations and Data Reduction}

\subsection{VLBA Observations}
The 1720 MHz transition of the ground-state OH molecule was observed
with the VLBA (Napier \etal 1994) on 09 May 1997 toward one $\sim$ 25\arcm 
field in each of the two supernova remnants \wte and \wffns.  
Table 1 lists the 
position of the pointing center for both sources.  These positions were
chosen to encompass both the ``OH E'' and ``OH F'' 1720 MHz masers in
\wte and the ``OH E'' masers in \wffns, following the nomenclature of Claussen 
\etal (1997).  A single antenna of the VLA was also used in conjunction 
with the antennas of the VLBA in order to provide short projected baselines
($\sim$60 km).  

The data were recorded with a 62.5 kHz bandwidth centered at
velocities of 44.0\kmss and 10.0\kmss 
(LSR) for \wff and \wtens, respectively.  
Both right and left circular polarizations were recorded
with 2-bit sampling. The data were correlated with the NRAO VLBA correlator 
to provide 128 spectral channels for each polarization 
averaged every 4.1 seconds.  All four polarization correlations
were performed.  This correlator mode provided a channel spacing of 
0.09\kmss per spectral channel. The velocity resolution is slightly
large than this ($\sim$0.11\kms) due to the spectral weighting function 
applied.

In order to provide manageable dataset sizes, the correlator averaging time  
was the limiting factor for the field of view.  For averaging times of 4.1 
seconds, the fringe-rate window for the longest baselines of the VLBA provides 
a field of view of approximately 24\arcs.  Thus for both SNRs we made two
correlation passes near positions of strong OH(1720 MHz) masers in the
primary field of view.  Table 2 provides the positions 
of the two correlation positions for both remnants, and the peak flux density in
the VLA {\bf A} configuration observations (Claussen \etal 1997).

Data reduction was performed using standard software 
contained in the NRAO AIPS package.  Delays were measured via observations of 
nearby continuum sources (1748$-$253 for \wtens, and 1904+013
for \wffns) by performing fringe fitting.  Bandpass
calibration was determined using total-power observations of the strong
sources 1921$-$293 and 1611+343.
Residual fringe rates were determined by fringe
fitting on a single strong spectral channel.  For \wtens, amplitude
calibration was accomplished by fitting the bandpass-corrected, 
total-power, on-source spectra of each antenna in the array to a template 
total-power, on-source spectrum observed with a sensitive antenna (at
the Mauna Kea, HI station) at a high elevation angle.  The absolute 
amplitude calibration determined by this method is accurate to about 10\%.  
For \wffns, the amplitude calibration was determined by using known antenna 
gains (provided by NRAO staff) and system temperatures measured during the 
observations (so-called {\it a priori} amplitude calibration).  This procedure 
was carried out for the \wff data because the signal-to-noise ratio in the 
total-power spectra of the OH maser line was not high enough to attempt the 
template-fitting method.  The absolute amplitude calibration determined by the 
{\it a priori} method is accurate to about 15\%.  Even on the strongest spectral 
channels, no correlated signal was detected on the longer baselines (typically 
from one of the southwest locations to Saint Croix, VI; Hancock, NH; and 
Mauna Kea, HI), and so data to these stations were automatically deleted.

The spectral channel with the greatest flux density was 
then used in an iterative self-calibration mapping procedure. These 
self-calibration solutions were then applied to all spectral channels. 
The rms noise ($\approx$ 100 mJy beam$^{-1}$)
in these images is close to the expected  theoretical noise limit.  Images
were then made of all spectral channels with maser emission.  Naturally 
weighted maps for each 
velocity channel were then produced using the AIPS task IMAGR.  
The resultant synthesized beam was about 40 $\times$ 15 mas
at a position angle of 0\deg for \wtens, and about 40 $\times$ 30 mas at
a position angle of $-$10\deg for \wffns. 

\subsection{MERLIN Observations}
The MERLIN radio telescope was used on 12 April 1998 to observe the
position listed in Table 1 for \wff in the 1720 MHz transition
of OH.  Seven telescopes were used, including the 76-m Lovell telescope,
the Mark II, the 32-m telescope at Cambridge, and the 25-m telescopes
at Tabley, Darnhall, Knockin, and Defford.  Both right and
left-hand polarization was observed, and the correlator produced
512 spectral channels over a bandwidth of 500 kHz, for a channel
spacing of 0.18 \kms.  The field of view for the MERLIN observations
included both the \wff ``E'' and ``F'' sources (Claussen \etal 1997).
The synthesized beam from the MERLIN observations was 290 $\times$ 165
mas at a position angle of 25\deg.

Initial phase and amplitude calibration 
was performed by MERLIN staff at the University of Manchester. 
Bandpass calibration was determined from observations of 3C~84, and 
phase calibration was determined from interleaved observations of
1904+013.  The absolute amplitude calibration was determined by 
observations of 3C~286 on the shortest baselines.

Similar to the VLBA reduction, AIPS was then used to apply 
self-calibration solutions, based on iterative self-calibration
imaging carried out on the strongest spectral channel.  The rms noise 
obtained after applying the self-calibration to a single channel
was about 30 mJy beam$^{-1}$, also close to the theoretical noise limit.  

\section{Results}

\subsection{\wte Masers}
Figure 1 shows a contour image of the OH(1720 MHz) emission at 
11.3\kmss, and Stokes {\it I} spectra at two emission peaks.
This emission corresponds to feature F~39 in the nomenclature of Claussen
\etal (1997) (see their Table 2).  The peak flux density in the image is
6.1 Jy beam$^{-1}$ with a total flux of $\sim$70 Jy, in close agreement with 
the peak flux density (73 Jy beam$^{-1}$) at this position in the VLA {\bf A} 
configuration maps.  The OH(1720 MHz) emission is clearly resolved at this 
resolution.  Several emission peaks can be seen. The size of the feature 
marked {\bf B} in Figure 1, determined by Gaussian fitting, is 
75 $\times$ 34 mas, at a position angle of 9\deg.  Thus the peak brightness 
temperature is $\sim$2 $\times$ 10$^9$ K.  Other emission peaks in 
the contour image of Figure 1 have similar spectral profiles, differing 
primarily in their peak flux density.  
In addition to the masers shown in Figure 1,  there is an additional maser
emission peak $\sim$500 mas to the northeast, at a velocity of 
9.6\kmss.  Figure 2 shows a contour image of the peak velocity channel in 
this region and Stokes {\it I} spectra of two emission peaks.  Again, the 
emission is quite extended (size $\sim$60 mas) compared with the beam.  
The two peaks {\bf C} and {\bf D} in the emission correspond to brightness 
temperatures of 1$\times$ 10$^9$ and 8 $\times$ 10$^8$ K, respectively.  

For the positions marked {\bf A} and {\bf B} in Figure 1, we have determined 
the Stokes parameters {\it I} and {\it V}.  Assuming the {\it V} profile is due
to the Zeeman effect, and that the splitting is small compared to the
intrinsic (Doppler) line width, the Stokes {\it V} profile  is proportional
to the frequency derivative of the Stokes {\it I} profile.  A fit of the 
{\it V} profile to the derivative of the {\it I} profile yields a 
measurement of the line-of-sight magnetic field (see Claussen \etal 1997
for further discussion).  Figure 3 shows the result of this fit for the
two positions marked in Figure 1.  The estimate of the line-of-sight magnetic 
field ($\sim$ 2 milliGauss) is larger by a factor of $\sim$10 than estimates based 
on the VLA observations.  However, toward this specific position, Claussen
\etal (1997) were unable to estimate the line-of-sight magnetic field because
the spectra were quite complex and did not show a clear signature (the
classical S-shape in Stokes {\it V}) of Zeeman splitting.  

This relation used to derive the line-of-sight magnetic
field is valid only for thermal absorption and emission lines.  Nedoluha
\& Watson (1992) conclude that the standard thermal relationship used here
is a valid approximation of the line-of-sight field strength for
observations of water masers, if they are not strongly saturated, despite
the complications of the maser radiative transfer.  Elitzur (1996, 1998)
has derived a general polarization solution for maser emission and
arbitrary Zeeman splitting.  According to this solution, Elitzur (1998) 
concludes that masers require smaller fields to produce the same
amount of circular polarization as thermal emission.  Thus the estimate 
of the magnetic field given above may be overestimated by as much as
factors of 2---4.

At the other correlated position in \wte (\wte E), we did not detect OH 
emission.  Based on the VLA observations, we expected that there should be
several features detectable by the VLBA.   Some of the flux densities 
measured in the VLA observations are 24 Jy/beam (E~30), 12.5 Jy/beam (E~31), 
and 10 Jy/beam (E~24).  E~24 was {\it unresolved} with the VLA.  If these
non-detections are due to emission that is smooth, we can calculate
a lower limit to the size of the emission features, based on the rms noise
that we measured on the shortest projected baseline in the VLBA observations.
Assuming a source size that is a circular Gaussian function, a maximum
visibility of 0.1 (for E~24, for example) on the shortest baseline
($\sim$60 km), the lower limit must be about 400 mas.  For E~30, the 
visibility must be correspondingly lower and thus the size must be larger
than about 500 mas.
\subsection{\wff Masers}
The OH(1720 MHz) masers observed with MERLIN are shown in Figures 4 and 5.
Figure 4 is a contour plot of the peak maser emission from the E~11 source
along with the OH spectrum of the peak emission.  The peak occurs at a 
velocity of 44.2\kmss.  The stronger of the two features in the contour image
has a peak flux density of 3.3 Jy beam$^{-1}$, and is slightly resolved with a 
fitted Gaussian size of 165 $\times$ 57 mas at a position angle of 147\deg.  
The  total flux density over the emission region is about 6.1 Jy, which is 
comparable to the peak flux density measured with the VLA.  The brightness 
temperature for this feature is  1.3 $\times$ 10$^8$ K.  We convolved the  
MERLIN map of the E~11 source with the VLA {\bf A} configuration beam, and 
then made a Gaussian fit to the resulting image.  The peak of the convolved 
image was 5.0 Jy beam$^{-1}$ with a fitted size of 715 $\times$ 250 mas at a 
position angle of 166\deg.  This is comparable with the VLA observations which 
obtained a peak flux density of 6.6 Jy beam$^{-1}$, and a fitted size of 890 
$\times$ 180 at a position angle of 137\deg. 

Figure 5 shows the OH(1720 MHz) maser emission from the \wff F~24 source
which peaks at a velocity of 46.9\kms, with a peak flux density of 1.5 Jy
beam$^{-1}$.  The OH spectrum at the emission peak is also shown.  The 
total flux density in the emission region is about 4.2 Jy, only about half 
of the VLA observed peak flux density.  

Figure 6 shows a contour plot of the VLBA image of the E~11 source.  This
maser source is unresolved at the resolution of the VLBA (40 $\times$ 30 mas).
The peak flux density is 1.2 Jy beam$^{-1}$, and thus a lower limit to the
brightness temperature is 8 $\times$ 10$^8$ K.  This source is the core
of the brightest MERLIN source shown in Figure 4.  If a single Gaussian
component with a large size were responsible for the difference in flux 
density between the VLBA and MERLIN measurements, then, based on the 
shortest projected spacing of the VLBA and the measured noise, the 
size of such feature would have to be larger than 270 mas.  This
is inconsistent with the MERLIN measurement.  Thus we conclude that the 
2.1 Jy of missing flux density must be in a few components whose peaks are 
each weaker than about 0.4 Jy beam$^{-1}$.

\section{The OH Maser Sizes: Scattering Disks or Physical Size?}
 
Table 3 summarizes the measured components and the estimated
brightness temperatures of the OH(1720 MHz) masers in both \wte and
\wffns.  The VLBA and MERLIN observations clearly resolve the masers
seen by Claussen et al. (1997) into multiple components. Typical
angular sizes in both SNRs are 50 to 100 mas with aspect ratios of
order 2.5:1. While these measurements have shown that the Claussen et
al. (1997) maser sizes were an artifact of the low resolution
used, a major question remains:  ``are these compact and 
resolved features the true size of the masers or are they due to
interstellar scattering ?''.   The observations and results of the present 
study cannot distinguish between these two options.
This is due mainly to our ignorance of what the intrinsic sizes might be.  
In what follows we will interpret the results of these observations 
as they apply to both intrinsic structure and that due to scattering,
and suggest further observational tests that should help to distinguish
between these two possibilities. 

\subsection{Scattering Interpretation}

If the maser sizes in \wff and \wte are dominated by scattering, other
nearby (in projection) objects such as pulsars and extragalactic
sources can also be affected.  
Pulse broadening and angular broadening depend in different ways upon 
the distribution of scattering material along the line of sight; 
additionally, angular broadening of an extragalactic source seen 
through the turbulence in our Galaxy's ISM is sensitive only to the 
strength of the turbulence, not to its distribution along the line of 
sight.  The relative distances of the masers and nearby pulsars along 
with the distribution of scattering material can all be constrained using
measurements of angular and pulse broadening.

Adopting a distance of 3 kpc for both \wte and \wff, the models of
Cordes et al. (1991) and Taylor \& Cordes (1993) predict angularly
broadened sizes for the OH masers of order 1$-$3 mas. These angular
broadening estimates are well below the sizes given in Table 3 but it
should be noted that the angular broadening is underestimated 
for lines-of-sight subject to enhanced scattering.  
Examples of lines-of-sight with enhanced scattering include the Galactic 
Center region (van Langevelde et al. 1992), the Cygnus region (Fey, 
Spangler \& Cordes 1991), and
towards 1849+005 (Fey, Spangler \& Cordes 1991).  We discuss separately the 
implications of our maser size measurements for the scattering towards 
\wte and \wffns.

\subsubsection{Scattering in the Direction of \wte}

The \wte SNR and its associated masers lie at a distance of
approximately 3 kpc (\eg Kaspi \etal 1993; Frail, Kulkarni,
\& Vasisht 1993) in the direction $(l,b)=(6.8,-0.06)$.  The 60,000
year old pulsar PSR~B1578$-$23 also lies in the same direction but is
located outside the SNR, a few arcminutes to the north of its bright
continuum edge.  An extragalactic source, 1758$-$231, lies within two
arcminutes of this pulsar.  Frail \etal (1993), using
observations of the pulsar and the neighboring extragalactic source,
argued in favor of the association of the pulsar and the SNR. 
Kaspi et al. (1993) disagreed, suggesting that the pulsar was much 
further away.  The discussion of Frail \etal was based upon a pulse 
broadening measurement of 70 milliseconds at 1~GHz for PSR~B1758$-$23 
(Kaspi \etal 1993) and upon an upper limit of $1$~arcsecond to the size 
of the extragalactic source 1758$-$231.  Here we review the implications of
our measurements of the $60$~mas maser size as it pertains to this
disagreement.

Under the usual assumption that masers and the extragalactic
source are scattered by a turbulence confined to a single thin screen,
angular broadening measurements completely constrain the location of
the screen.  The two angular broadening sizes are related to the
location of the screen by $$d_s/d_m = 1-\theta_m/\theta_e ,$$ where
$\theta_m$ and $\theta_e$ are the angular broadening sizes of the
masers and the extragalactic source and $d_m$ and $d_s$ are the 
observer distances to the masers and the screen, respectively.  
If the masers and the extragalactic source have sizes of 
$\theta_e = 1$\arcsec and $\theta_m = 0.06$\arcsec, respectively, then
for a maser distance of $d_m=3$~kpc, the scattering screen lies
at $d_s = 2.8$~kpc.
Note that the screen distance decreases if $\theta_m/\theta_e$ increases.

The pulse broadening of PSR1758$-$231 and the angular broadening of
1758$-$231 provide a general constraint 
on the location of the pulsar.  Frail \etal (1993) assumed a distance
of 3~kpc for PSR~B1758$-$23 and combined the pulse broadening of
PSR~B1758$-$23 with the angular broadening of 1758$-$231 to constrain
$f_p$, the ratio of the observer-screen distance to the screen-pulsar
distance, to values between $0.3$ and $3.4$.  
A more general constraint on the location of the pulsar can be obtained
if the equations presented by Frail \etal are combined without assuming
a distance to the pulsar to produce
$$\theta_e = {0.42\over\sqrt{d_p/3}}
\left(\sqrt{f_p}+{1\over\sqrt{f_p}}\right) ,$$ where $\theta_e$ is the angular size
of the extragalactic source in arcseconds, and $d_p$ is the distance
from the observer to the pulsar in kpc.  It is easy to show that this
implies that the distance to the pulsar is given by $$d_p = 2.12 /
\theta_e^2 ,$$ if the screen is halfway to the pulsar and is further
otherwise.  In particular, if the measured extragalactic source size 
is smaller than $0.84$~arcseconds, the pulsar cannot be associated with 
the \wte SNR.

\subsubsection{Scattering in the Direction of \wff} 

The \wff SNR and its associated masers lie at a distance of
approximately $3$~kpc (Radhakrishnan \etal 1972) in the direction 
$(l,b)=(34.7, -0.4)$.  The
association of pulsar PSR~B1853$+$01 with \wff is well established
based on ages, dispersion measure distance, and positional coincidence
(Wolszczan, Cordes, \& Dewey 1991).  Assuming a $3$~kpc distance to 
PSR~B1853$+$01, the
Taylor \& Cordes model predicts $30$~microseconds of pulse broadening
and $1$~mas of angular broadening.  Because the Taylor \& Cordes
model prediction of $92$~mas severely underestimates the observed 
angular broadening size of $378$~mas for 1849+005, an extragalactic 
source situated only $3.7$ degrees away from \wffns, we consider the 
possibility that the observed $100$~mas OH maser sizes are due to 
angular broadening.  The observational limit to the pulse broadening 
is of order $1$ millisecond.

Lines-of-sight with enhanced scattering are believed to intersect a
clumped component of scattering material in the interstellar medium.
All lines-of-sight passing within $30$~arcminutes of the Galactic
Center are known to be heavily scattered (van Langevelde \etal 1992;
Lazio \& Cordes 1998).
The scattering material has been constrained to lie within 50 parsecs
of the Galactic Center.  At a distance of $8.5$~kpc, a clump of size
$30$~arcminutes corresponds to enhanced scattering in a region of 
about $40$~pc.  Similar clump sizes have been proposed for scattering
towards other lines-of-sight (\eg Dennison \etal 1984).

Because the OH masers in \wff and PSR~B1853$+$01 both lie at $3$~kpc,
the distance $d_s$(in kpc) to an assumed scattering screen can be derived from
the pulsar size formula presented by Frail \etal: 
$$d_s = {3\over1+{\theta_m^2/2.52\tau}} ,$$
where $\theta_m$ is measured in arcseconds and $\tau$, the pulse broadening,
is measured in seconds.  If the angular broadening size for OH masers in 
\wff is assumed to be $80$~mas and the pulse broadening of order 
$0.5$~milliseconds, a
clump of enhanced scattering would need to be placed only $0.5$~kpc 
away.  At this distance, a $40$~pc clump would subtend an angle of 
about $4$ degrees and could also intercept the line of sight towards 1849+005.
Since the line-of-sight towards 1849+005 is the second most heavily
angularly broadened line of sight known, it is reasonable to suggest
that other, neighboring, lines-of-sight should also be heavily
scattered --- as is the case for the Galactic Center direction.  The
observed sizes of OH masers in \wffns, if dominated by angular
broadening, have sizes consistent with the scattering of 1849+005.
This hypothesis predicts that other lines-of-sight close to 1849+005
should also be heavily scattered.  In addition, improved better measurements of the
pulse-broadening towards PSR~B1853$+$01 could help to prove or
disprove this suggestion.

\subsection{Intrinsic Structure of the Masers}

If the structure that is observed in the OH(1720MHz) masers toward
\wff and especially \wte are due to variations in emission intrinsic 
to the maser, then this data is the first demonstration of structure in
1720 MHz OH maser emission.  The pumping requirements of these
OH (1720 MHz) masers, modeled by Lockett \etal (1999), strongly suggests
an OH column density of 10$^{16}$ cm$^{-2}$ with molecular
hydrogen density $\sim$ 10$^5$ cm$^{-3}$. This requires a linear
dimension of ${10^{11}} \over {x_{OH}}$ cm, where $x_{OH}$ is the OH abundance.
According to Lockett \etal (1999), the highest OH abundance expected in 
C-shocks is $\sim 2\times10^{-5}$, so the expected thickness of the OH
emitting region is about 5$\times10^{15}$ cm, similar to the maser sizes
we observe.  

We could conclude that these masers appear to be similar to the stellar 
OH(1612 MHz) masers.  The OH(1612 MHz) masers in circumstellar shells 
show a large range of size scales: 40 ---
1000 mas, as demonstrated by Bowers \etal (1990).  Both main-line
and 1720 MHz OH found in the interstellar medium have been shown to
be very compact and with little structure (\eg Reid \etal 1980;
Forster \etal 1982; Masheder \etal 1994).  Bowers \etal (1990) suggest 
that the OH(1612 MHz) maser structure is determined by a combination of 
density and velocity effects; our 1720 MHz observations could be indicative 
of a similar situation.

A good test of whether or not the OH(1720 MHz) emission is really due to 
intrinsic structure would be a high-resolution observation of the OH(1720 MHz)
emission toward a nearby SNR in the anti-center direction (to minimize
possible scattering effects).  A good candidate for this test 
observation would be the SNR IC~443.  It lies at a Galactic longitude
$\approx189$\deg, and is only 1.5 kpc distant.  If a measurement of the
size of the OH(1720 MHz) masers in IC~443 showed the masers to be smaller than
those in \wte or \wffns, then a good argument for scattering of the masers in 
\wte and \wff could be made.  If the sizes of the masers in IC~443 were 
similar in size or larger than those in \wte or \wff, then the sizes of
the masers in all three of the SNR would likely be intrinsic.

\section {The OH Maser Polarization in \wtens:  Magnetic Fields}
The measurement of the line-of-sight magnetic field of about 2 
milliGauss toward the region of strong maser emission in \wte 
is stronger than the more widespread measurements reported by 
Claussen \etal (1997) toward the SNR by about a factor of 10.
This measurement is closer to that reported by Yusef-Zadeh 
\etal (1996, 1998) for magnetic fields in the Galactic Center.
It is interesting to note that the strongest field strength
we measure is also in the region of strongest maser emission.
This may be a selection effect, since our observations were
limited to only a few maser regions.  Since both the current
observations and the VLA observations of {\it V} profiles show
the classical S shapes of the Zeeman effect, we are confident
that both sets of measurements are good estimates of the 
line-of-sight magnetic field.

In this small region, the magnetic pressure must be 100 times
the pressure estimated by Claussen \etal (1997), or about 
2$\times$10$^{-7}$ dyn cm$^{-2}$.  Thus the magnetic pressure
is very much larger than the thermal gas pressure of 6$\times$
10$^{-10}$ dyn cm$^{-2}$ estimated from hot X-ray gas in the
interior of the remnants (Rho \etal 1996), and so the magnetic
field is likely the dominant factor in the structure of the
shock.  As discussed by Lockett \etal (1999) and Draine, 
Roberge, and Dalgarno (1983), the larger magnetic field estimated
here is further strong evidence for a C-type shock in the OH
maser region.

\section{Conclusions}

We have used the VLBA and MERLIN to observe some of the OH(1720 MHz) 
masers toward the two supernova remnants \wff and \wte at resolutions 
of 40 mas.  We have resolved the masers in both SNRs.  The range of 
observed sizes are 50$-$180 mas, and the derived apparent brightness 
temperatures are in the range 0.3$-$20$\times{10}^8$ K.  Based on the 
present data, it is unclear if the observed structure of the masers is 
due to interstellar scattering or to intrinsic structure.

If the OH(1720 MHz)structure is due to intrinsic maser emission, then we 
suggest that the OH(1720 MHz) masers in SNR may be similar to the OH(1612 MHz) 
maser emission from circumstellar shells.  A possible test of whether or not 
the size and structures observed are intrinsic or due to scattering would be a
high angular resolution observation of the remnant IC~443.

If the sizes measured are considered to be dominated by the angular 
broadening effects of interstellar scattering, conclusions can be 
drawn about the location of the scattering material.  In the case of \wtens,
the scattering material along the line of sight is most likely 
situated within $\sim100$~pc of the masers.  Also, we would conclude that 
pulsar PSR~B1758$-$23 is definitely {\it not} associated with the SNR.  
In the case of \wffns, we suggest that a 40~pc clump of enhanced 
scattering material located 500 pc from the Sun could explain the 
observed maser sizes as well as the scattering of 1849$+$005.
This suggestion could be tested by searching for other heavily scattered 
sources within $\approx4$ degrees of 1849+005 and by improved pulse
broadening measurements of the pulsar PSR~B1853$+$01.

Finally, we measure a magnetic field in a small region of the SNR 
\wte ($\sim$2$\times$10$^{15}$ cm), which is a factor of
about 10 higher than that which was measured using the VLA.  The
magnetic field clearly dominates the shock structure, and is
further evidence for a C-type shock in the OH maser region.

\acknowledgments
The National Radio Astronomy Observatory is a facility of the National 
Science Foundation, operated under cooperative agreement by Associated 
Universities, Inc. MERLIN is a UK national facility operated by 
the University of Manchester on behalf of PPARC. We thank Peter Wilkinson
for granting us time on MERLIN from the Director's discretion, Peter 
Thomasson for his invaluable assistance in scheduling MERLIN, and Anita
Richards for performing the initial phase and amplitude calibration
for the MERLIN data.  Finally, we thank the referee, Moshe Elitzur, for
useful comments and discussions which have improved the paper.

\clearpage
\begin{table}
\begin{center} 
\caption{Pointing Positions of VLBA Observations}
\begin{tabular}{lcc}
Source & RA (J2000) & Decl. (J2000) \\
\tableline
\tableline
W~28 & 18$^h$ 01$^m$ 51$^s$.473 & $-$23$^o$ 17\arcmin 49\arcsec.64 \\
W~44 & 18$^h$ 56$^m$ 29$^s$.099 & $+$01$^o$ 29\arcmin 43\arcsec.36  \\
\tableline
\end{tabular}
\end{center}
\end{table}
\begin{table}
\begin{center}
\caption{VLBA Correlation Positions and VLA Peak Flux Densities}
\begin{tabular}{lccc}
Source & RA (J2000) & Decl. (J2000) & Peak Flux Density (Jy beam$^{-1}$  \\
\tableline
\tableline
W~28 OH E & 18$^h$ 01$^m$ 51$^s$.198 & $-$23$^o$ 17\arcmin 40\arcsec.64 & 24.4  \\
W~28 OH F & 18$^h$ 01$^m$ 52$^s$.706 & $-$23$^o$ 19\arcmin 24\arcsec.64 & 73.0  \\
W~44 OH E & 18$^h$ 56$^m$ 26$^s$.665 & $+$01$^o$ 29\arcmin 43\arcsec.36 &  6.6  \\
W~44 OH F & 18$^h$ 56$^m$ 36$^s$.627 & $+$01$^o$ 26\arcmin 35\arcsec.89 &  8.9  \\
\tableline
\end{tabular}
\end{center}
\end{table}

\begin{table}
\begin{center}
\caption{Measured Properties of 1720 MHz OH Masers in \wte and \wff}
\begin{tabular}{lccccc}
Source & Major Axis & Minor Axis & P.A. & Peak Flux Density & T$_B$ \\
& (mas) & (mas) & (deg) & (Jy beam$^{-1}$) & (10$^8$ K) \\
\tableline
\tableline
W~28 F~39 A & 105 & 34 & 20 & 5.5 & 10  \\
W~28 F~39 B & 75 & 34 & 9 & 6.1 & 20  \\
W~28 F~39 C & 68 & 39 & 21 & 3.2 & 8  \\
W~28 F~39 D & 58 & 31 & 24 & 3.1 & 10  \\
W~44 F~24 & 240 & 135 & 121 & 1.5 & 0.3  \\
W~44 E~11 (MERLIN) & 165 & 57 & 147 & 3.3 & 2  \\
W~44 E~11 (VLBA) & $<$40 & $<$30 & --- & 1.2 & $>$6  \\
\tableline
\end{tabular}
\end{center}
\end{table}
\clearpage
\begin{figure}
\caption{A VLBA contour image of the OH(1720 MHz) maser emission from feature
F~39 in \wtens, following nomenclature of Claussen \etal (1997).  The contours 
are plotted at 0.3, 0.53, 0.93, 1.63, 2.86, and 5.00 Jy beam$^{-1}$.  The insets show
Stokes {\it I} spectra taken at the peaks marked {\bf A} and {\bf B} by 
crosses.}
\end{figure}
\begin{figure}
\caption{A VLBA contour image of the OH(1720 MHz) maser emission in \wtens, 
about 500 mas to the northeast of the emission shown in Figure 1.  The contours
are plotted at the same levels as in Figure 1.  The insets show Stokes 
{\it I} spectra taken at the emission peaks marked {\bf C} and {\bf D}
by crosses.}
\end{figure}
\begin{figure}
\caption{(a) For the \wte position marked {\bf B} from Figure 1, the Stokes {\it V}
profile (solid line) and the derivative of the Stokes {\it I} profile,
scaled by $+$2.2 milliGauss (line-of-sight field). (b) As (a), but for the 
\wte position marked {\bf A} from Figure 1.  The scaling here is by $+$1.8 milliGauss 
(line-of-sight field).}
\end{figure}
\begin{figure}
\caption{A MERLIN contour image of the OH(1720 MHz) maser emission from 
the \wff E~11 source.  The contours are plotted at $-$0.05, 0.05, 0.1,
0.2, 0.4, 0.8, 1.6, 3.2 Jy beam$^{-1}$.  The inset shows an OH spectrum
taken at the peak in the contour image.}
\end{figure}
\begin{figure}
\caption{A MERLIN contour image of the OH(1720 MHz) maser emission from 
the \wff F~24 source.  The contours are plotted at $-$0.075, 0.075, 0.15,
0.3, 0.6, 1.2 Jy beam$^{-1}$.  The inset shows an OH spectrum
taken at the peak (marked by a cross) in the contour image.}
\end{figure}
\begin{figure}
\caption{The VLBA contour image of the OH(1720 MHz) maser emission from the
\wff E~11 source. }
\end{figure}
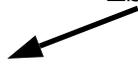
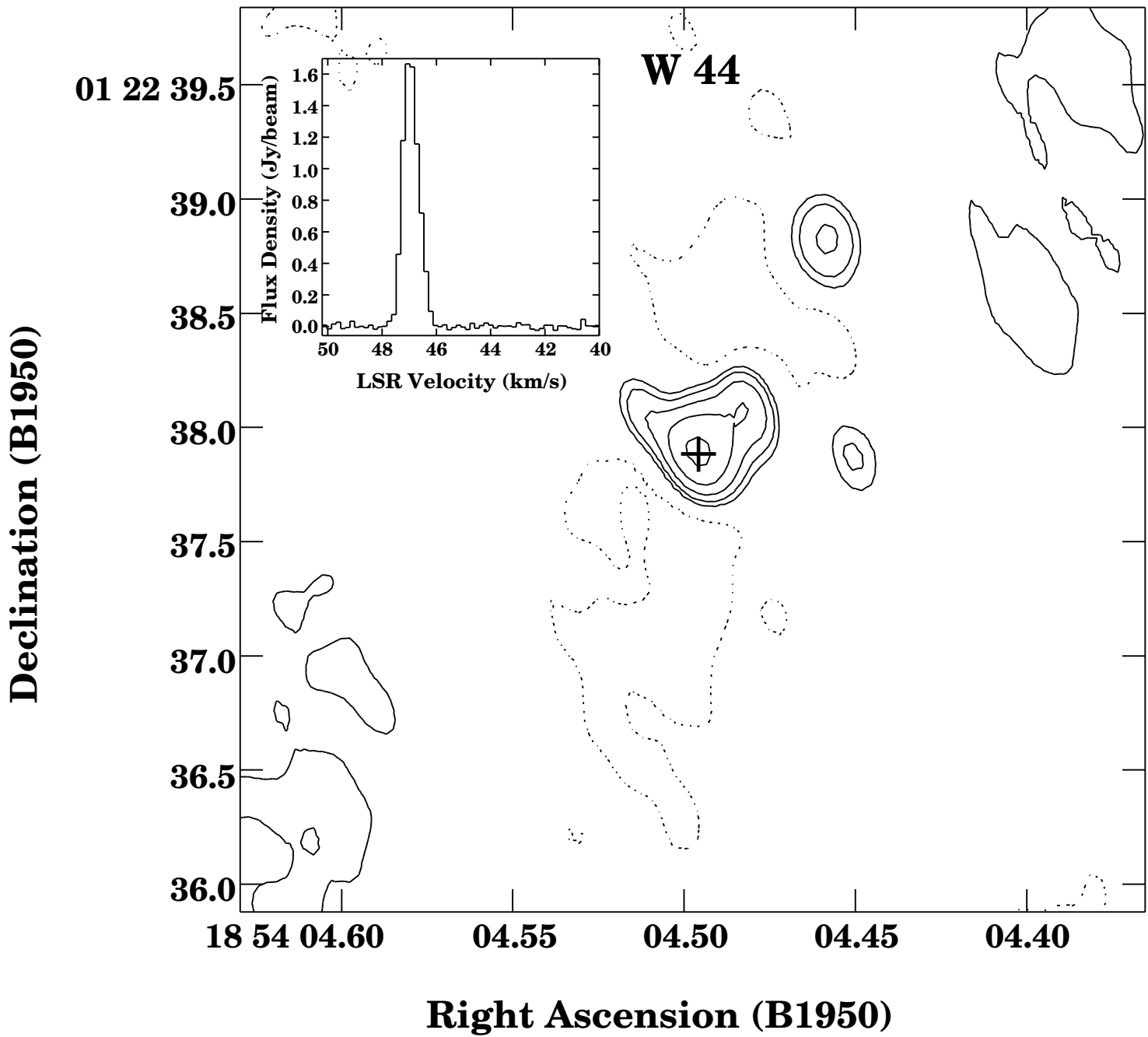
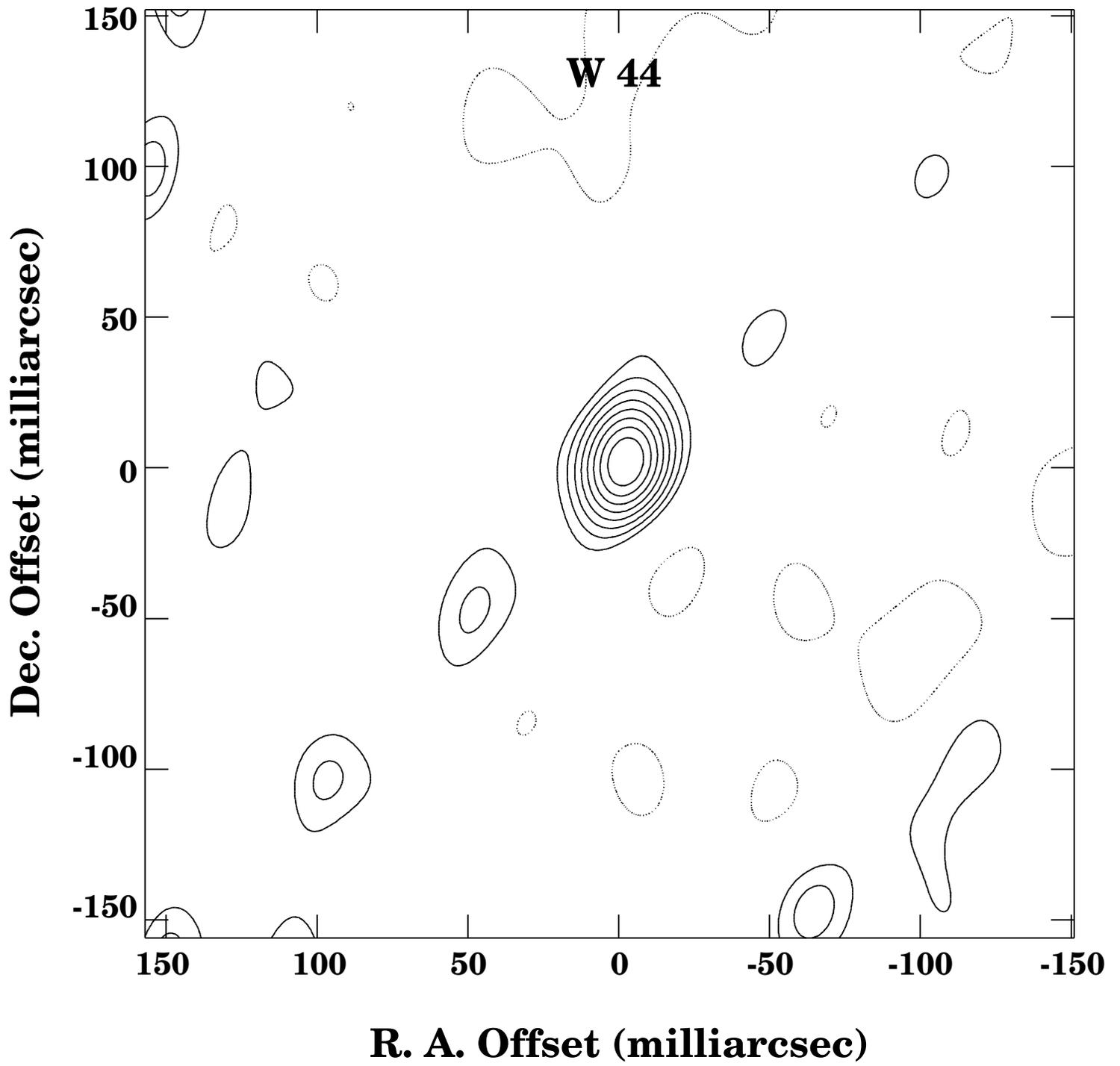
\end{document}